# Laser and pinching discharge plasmas spectral characteristics in water window region

P Kolar[1], M Vrbova[1], M Nevrkla[2], P Vrba[2,3] and A Jancarek[2]

[1]Czech Technical University in Prague, Faculty of Biomedical Engineering, Czech Republic
[2]Czech Technical University in Prague, Faculty of Nuclear Sciences and Physical Engineering, Czech Republic
[3]Institute of Plasma Physics, Academy of Sciences, Czech Republic

E-mail: petr.kolar@linet.cz

**Abstract**

Water window emission spectra of two laboratory nitrogen plasma sources were examined with a free standing grating spectrometer (FSGS). The highest line intensities at the wavelength 2.88 nm achievable with the sources were compared. Pulse energies for this line were judged as 0.02 mJ srad$^{-1}$ and 0.16 mJ srad$^{-1}$ with laser produced plasma system and pinching discharge plasma, respectively. The spectral resolution was found about 0.01 nm at 3rd order of spectral lines around wavelength 2.5 nm.

## 1. Introduction

Nitrogen plasma may be a suitable source of intensive line radiation in the water window spectral region (2.4–4.4 nm), if the plasma electron temperature is several tens of electronvolts. This intensive line radiation corresponding to the $1s^2 - 1s2p$ quantum transition of helium–like nitrogen ions may be useful namely for imaging of small biological objects in water environment.

There are several published papers related to radiation sources based on nitrogen plasma, designed to operate in water window region. Namely, laser produced nitrogen plasma systems: Fiedorowitz et al reached $6.2 \times 10^{11}$ photons srad$^{-1}$ by laser produced plasma system [1] and Wachulak et al $4.5 \times 10^{12}$ photons srad$^{-1}$ by laser system with double nozzle at wavelength 2.88 nm [2]. Peth reports in his dissertation [3] the value $5.0 \times 10^{11}$ photons srad$^{-1}$ reached by laser produced plasma system. Emitted energies $4.5 \times 10^{12}$ photons srad$^{-1}$ with capillary discharge nitrogen plasma are presented by Bergmann et al when pseudosparklike system was used [4]. Professsional electrodeless z-pinch device EQ-10M of company Energetiq is able to generate about $1.6 \times 10^{12}$ photons srad$^{-1}$ with repetition rate up to 1 kHz [5].

We have examined both types of nitrogen laboratory plasma sources, created either by neodymium YAG laser radiation focused to nitrogen gas jet or by pinching discharge in nitrogen filled capillary . Our spectral diagnostics was focused on the water window spectral range with predominating wavelength of 2.88 nm, corresponding to the $1s^2 - 1s2p$ quantum transition of helium–like nitrogen. The laser plasma laboratory radiation source was made by Laser-Laboratorium Göttingen (LLG) in Germany. In this system 500 mJ Nd:YAG laser pulse of 5 ns duration was focused by lens with the focal length 10 cm [3].

The system of pinching capillary discharge was developed at our University by Faculty of Nuclear Sciences and Physical Engineering in Prague (FNSPE) [6]. Alumina capillary of 3.2 mm diameter and 20 cm length is filled with nitrogen of tuned initial pressure 40 - 80 Pa. The 300 ns current pulse with amplitude 13,5 kA is created by discharging 25 nF capacitors initially charged to 70-80 kV.

## 2. Experimental setup

In our spectrometer we have used free-standing silicon nitride grating with period $d = 100$ nm and the ratio of the width of free space $a$ and the grating period $d$ estimated to 0.5. This value was stated according to the record from scanning electron microscope. The thickness of grating is approximately 200 nm. The bar profiles are more trapezoidal than rectangular. There is also a transversal support structure with period about 1.5 µm.

Diagnostics of both sources has been performed with the same spectrometer construction. Proportions of the spectrometer are in both cases $L_1$ = 0.98 m (source–grating), $L_2$ = 0.17 m (grating–CCD), $D$ = 1.15 m (total distance). CCD back iluminated XUV camera Rigaku X–Vision M25 was used. The field diffracted by the grating is recorded by CCD camera coupled with the metalic bellows, which enables side moving in direction of dispersion. The grating is placed in the axis of the arm rotation. Fig. 1 shows the scheme of the experimental setup of LLG system. LLG laser plasma shape is elongated in laser beam direction which lies in horizontal plane on axis of spectrometer. Hence, to maximize a spectral resolution the arm of spectrometer revolves in a vertical plane (Fig.1). A slit of width 30 µm was placed in front of the grating under the microscope to increase the spectral resolution.

The arrangement of measurements with the FNSPE capillary system was almost identical except aperture 2 (Fig. 2) selecting the central part of the output beam, because the plasma column is much wider then in the previous case.

Spectra were integrated over 120 shots with LLG system and 30 shots with FNSPE capillary sytem. The repetition rate used was 2 Hz for both measurements. The maximum available repetition rate of LLG source is 2 Hz, whereas the capillary source could achieve the repetition up to ten Hertz.

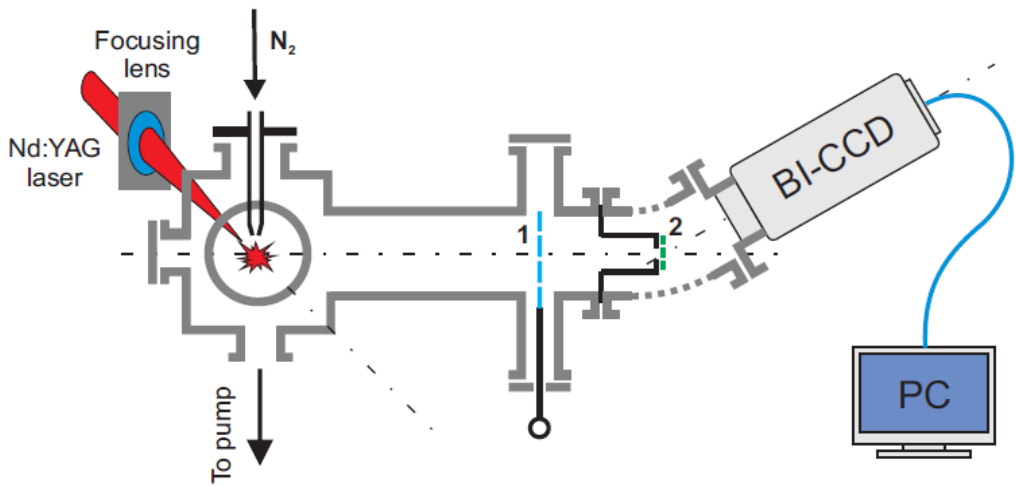

Fig. 1 *Experimental setup of LLG system spectral diagnostics measurement 1-Set of filters on the moving arm, 2-Free-standing diffraction grating*

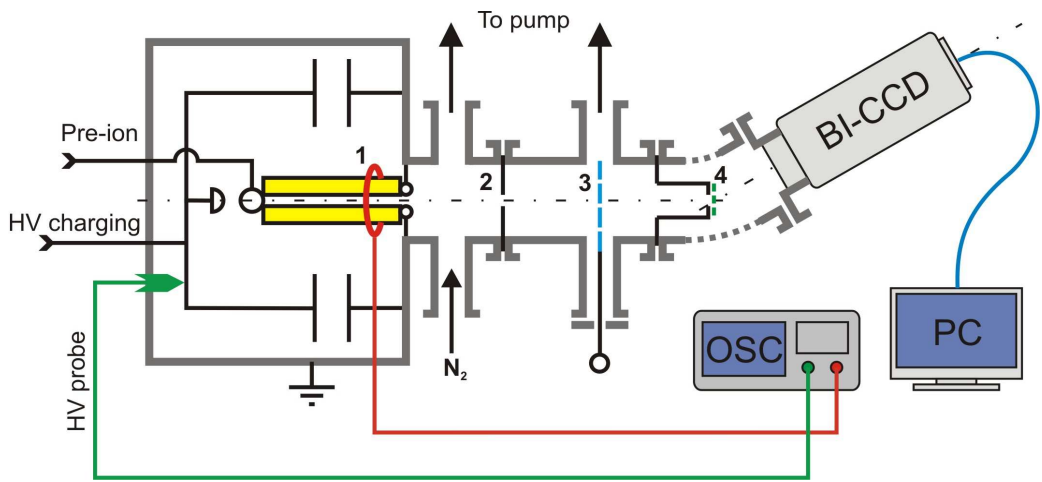

Fig. 2 *Experimental setup of FNSPE system spectral diagnostics measurement 1-Rogowski coil, 2-Aperture, 3-Set of filters on the moving arm, 4-Free-standing diffraction grating*

## 3. Results and discussions

Record of the diffraction pattern obtained with LLG system is shown in Fig. 3, where the most faraway very intensive lines belong to 1st order of 2.88 nm wavelength. There are also subsequent diffraction orders in the extension of this spectrum, where it is possible to reach resolution down to 0.01 nm due to higher dispersion. In the 3rd order of spectral lines 2,48 nm and 2,49 nm we can resolve transitions NVII 1s-2p, NVI 1s2-1s.3p respectively [7]. Although the free-standing grating seems to have ratio $a/d = 0.5$ which should eliminate even diffraction orders, there are evident lines of 2.88 nm radiation up to 6th order (Fig. 4 and 5). This occurs due to shape and transmission of silicon nitride grating bars.

Titanium filter transmission characteristics [8] (as seen from Fig. 4 - green dashed line) allows us to isolate quasi-monochromatic radiation of referenced sources. We present measured characteristics in wavelenth region of Ti filter maximum transmission (Fig. 4 and 5). Generaly LLG system with Ti filter produces radiation with good monochromaticity of wavelength 2.88 nm (Fig 4).

Fig. 5 shows measured spectrum of FNSPE capillary source. One can see other weak lines beside 2.88 nm line in Ti filter transmission region. These components belong to hydrogen-like and helium-like carbon ion transitions. Carbon ions originate probably from insulator oil surrounding the capillary capacitor bank. [6] There are found also a few peaks of radiation in the region about 2.5 nm in both cases.

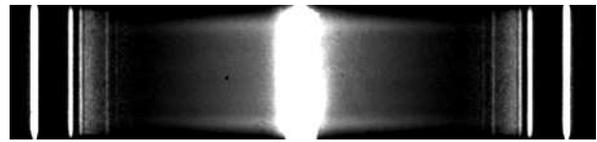

Fig. 3 *Record of LLG spectrum 0th order surrounding with most excentric line 2.88 nm, without filter, 120 pulses*

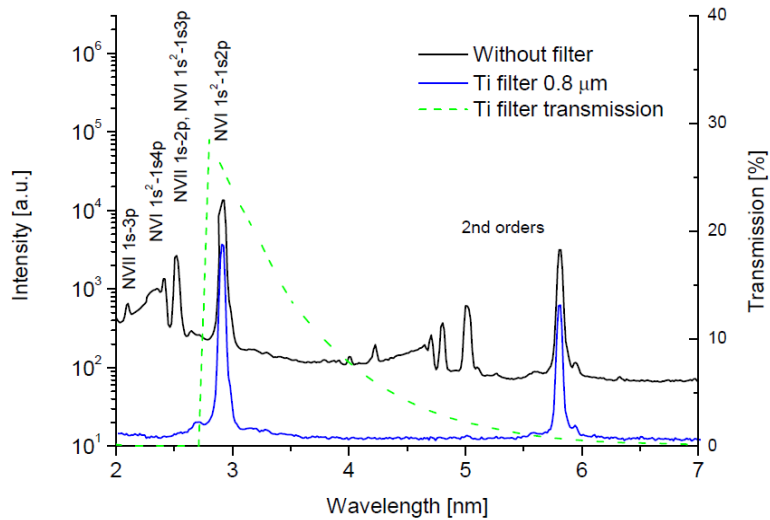

Fig. 4 *Recorded spectrum of LLG source (full black line), recorded spectrum with titanium filter (full blue line) and titanium filter transmission (green dashed line)*

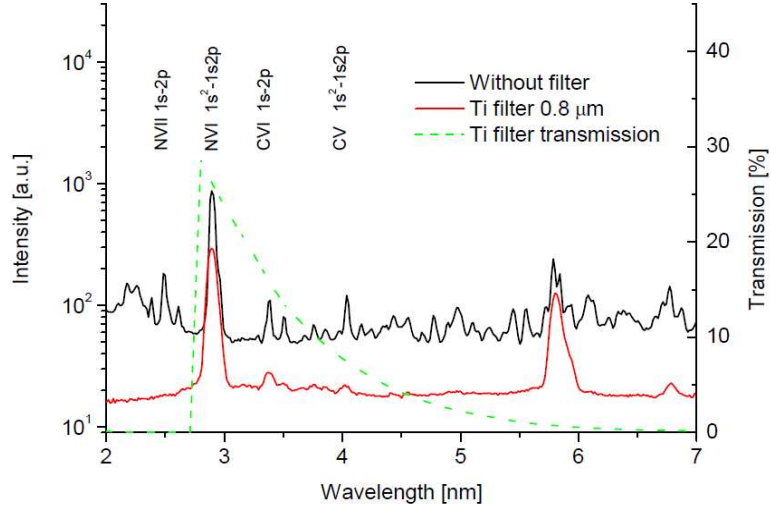

Fig. 5 *Recorded spectrum of FNSPE source (full black line), recorded spectrum with titanium filter (full red line) and titanium filter transmission (green dashed line)*

To estimate pulse energy of LLG radiation source two methods were used. The energy of the most intensive line transmitted trough a small aperture and a titanium filter was guessed from the CCD record. An aperture of 1 mm diameter covered with Ti filter supported by 5 µm nickel mesh was placed 0.2 m in front of a CCD camera. The distance between the source and the CCD was 1 m. Transmission of the 0.8 µm thick titanium filter about 30% was taken into account. Projection of the aperture was integrated over 20 pulses (Fig. 6) and then the mean value of signal intensity was evaluated. Energy of LLG system radiation per pulse was estimated as 0.01 mJ, which corresponds to $1.5 \times 10^{11}$ photons srad$^{-1}$. Energy judged from 1st diffraction order of radiation component 2.88 nm with using diffraction grating was 0.02 mJ srad$^{-1}$ and $2.8 \times 10^{11}$ photons srad$^{-1}$.

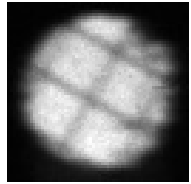

Fig. 6 *Projection of 1 mm dia. aperture covered with Ti foil supported by 5 µm nickel mesh*

Energy of FNSPE capillary source radiation per pulse was examined only with the grating method, which gives $2.3 \times 10^{12}$ photons srad$^{-1}$ and 0.16 mJ srad$^{-1}$.

The plasma temperature was estimated from the ratio of line intensities of transitions of the same line series presuming Boltzmann distribution of the corresponding energy level populations. Table 1 shows relative intensities of selected spectral lines. Transition $1s^2$–1s4p was not recognized in case of FNSPE source. Resulting plasma temperatures are estimated roughly 50-100 eV for both sources.

**Table 1.** *The intensities of selected spectral lines*

| System | Transition | Intensity [a.u.] |
|---|---|---|
| LLG | $1s^2$–1s2p | $10^4$ |
|  | $1s^2$–1s4p | $10^3$ |
|  | $1s^2$–1s5p | $8 \times 10^2$ |
| FNSPE | $1s^2$–1s2p | $4.5 \times 10^4$ |
|  | $1s^2$–1s5p | $5 \times 10^3$ |

Bandwidth due to Doppler broadening was calculated for the highest temperature of the interval and allows us to estimate ratio $\lambda/\Delta\lambda \approx 5 \times 10^3$. This value is expected as sufficient for microscope techniqe using Fresnel zone plates which, requires bandwith at least $\lambda/\Delta\lambda > 300$.

**5. Conclusion**

We have presented emission characteristics of two laboratory plasma sources in the water window spectral region. Both sources have strong lines in this region. In combination with the titanium filter they provide an almost monochromatic radiation with the wavelengts 2,88 nm and they are suitable for imaging experiments with biological samples containing water. The LLG laser plasma system emits approximately 2-3 $\times 10^{11}$ photons srad$^{-1}$ per pulse The radiation intensity of the FNSPE source is $2.3 \times 10^{12}$ photons srad$^{-1}$ e.i. nearly one order higher than that of the LLG system. The achievable repetition rate of FNSSPE system is also one order higher. The bandwidth due to Doppler broadening was evaluated for plasma temperature to $\lambda/\Delta\lambda \approx 5 \times 10^3$. As the FNSPE radiation source in the water window contains components of carbon ion transitions, the LLG system produces more monochromatic radiation.

## Acknowledgments

The research has been supported by the Ministry of Education Youth and Sports under the Development project C 42 - CTU 2010 and in the frame of INGO program on project No. LA08024.